\documentclass[proof]{WileyASNA-v1}

\usepackage{ulem}

\articletype{Article Type}%

\received{1 October 2020}
\revised{12 October 2020}
\accepted{25 October 2020}

\usepackage[latin1]{inputenc}

\newcommand{\beq}{\begin{equation}}
\newcommand{\eeq}{\end{equation}}
\newcommand{\ba}{\begin{array}}
\newcommand{\ea}{\end{array}}
\newcommand{\bea}{\begin{eqnarray}}
\newcommand{\eea}{\end{eqnarray}}
\newcommand{\bi}{\begin{itemize}}  
\newcommand{\ei}{\end{itemize}}
\newcommand{\ben}{\begin{enumerate}} 
\newcommand{\een}{\end{enumerate}}
\newcommand{\bc}{\begin{center}}
\newcommand{\ec}{\end{center}}

\newcommand{\De}{\Delta}
\newcommand{\ep}{\varepsilon}

\newcommand{\ptrans}{p_{\rm trans}}

\newcommand{\etrans}{\varepsilon_{\rm trans}}

\newcommand{\destab}{\De\ep_{\rm crit}}

\newcommand{\dd}{\textrm{d}}

\begin{document}

\title{On the energy budget of the transition of a neutron star into the third family branch}

\author{David E. Alvarez-Castillo*}

\authormark{David E. Alvarez-Castillo}

\address[]{\orgdiv{Henryk Niewodnicza\'nski}, \orgname{Institute of Nuclear Physics}, \orgaddress{\state{Cracow}, \country{Poland}}}

\corres{*ul. Radzikowskiego 152 31-342 Cracow, Poland.  \email{dalvarez@ifj.edu.pl}}

\abstract{Transition of a compact star into the third family for an equation of state (EoS) featuring mass twins is considered. The energy released at a baryon number conserving transition for static compact stars configurations is computed for two sets of models for comparison. The EoS of choice is the density dependent functional DD2 EoS with excluded model correction for hadronic matter which suffers a phase transition into deconfined quark matter described by a constant speed of sound approach. The two sets of EoS models feature different compact star mass onsets that maximize the energy and radius difference at the transition while simultaneously fulfilling state-of-the-art constraints from multi-messenger astronomy and empirical nuclear data. It is found that the maximal energy budget at the transitions falls in the range of $10^{49}$ - $10^{52}$ ergs. 
 }


\maketitle

\section{Introduction}

Neutron stars are extremely dense compact objects only surpassed by black holes. Determination of matter content in their cores has been a topic of intense research during the last years, with many developments achieved by constraints derived not only from laboratory experiments but also from astronomical observations. Out of the many possible realizations of the composition of neutron stars, the appearance of deconfined quark matter in their interiors turns out to be useful for probing the QCD phase diagram of dense nuclear matter, in which neutron star matter is located in the low temperature region. In addition, the \textit{mass twins} compact star phenomenon provides observational features worth of consideration. It consists of an EoS capable of producing two stars of about the same gravitational mass but different radius, located in the second and third branch of its sequence, as seen in a mass-radius diagram~(\cite{Glendenning:1998ag, Schertler:1998cs, Benic:2014jia,Alvarez-Castillo:2017qki,Montana:2018bkb}). For a comprehensive review on the various astrophysical aspects of mass twins, see~(\cite{Blaschke:2019tbh}).  

The purpose of this study is to estimate the maximum energy released in the transition of a neutron star into the third family branch for an appropriate equation of state (EoS) model that obeys state-of-the-art constraints derived from recent multi-messenger astronomy observations. This transition occurs once the central density of the compact star under consideration increases up to the critical value for quark matter deconfinement. Possible scenarios that can trigger this transition include either isolated neutron stars that spin down due to dipole radiation emission or neutron stars in binary systems that spin up by accretion of matter from the companion.

\begin{figure}[htpb!]
\center	
\includegraphics[width=70mm,height=50mm]{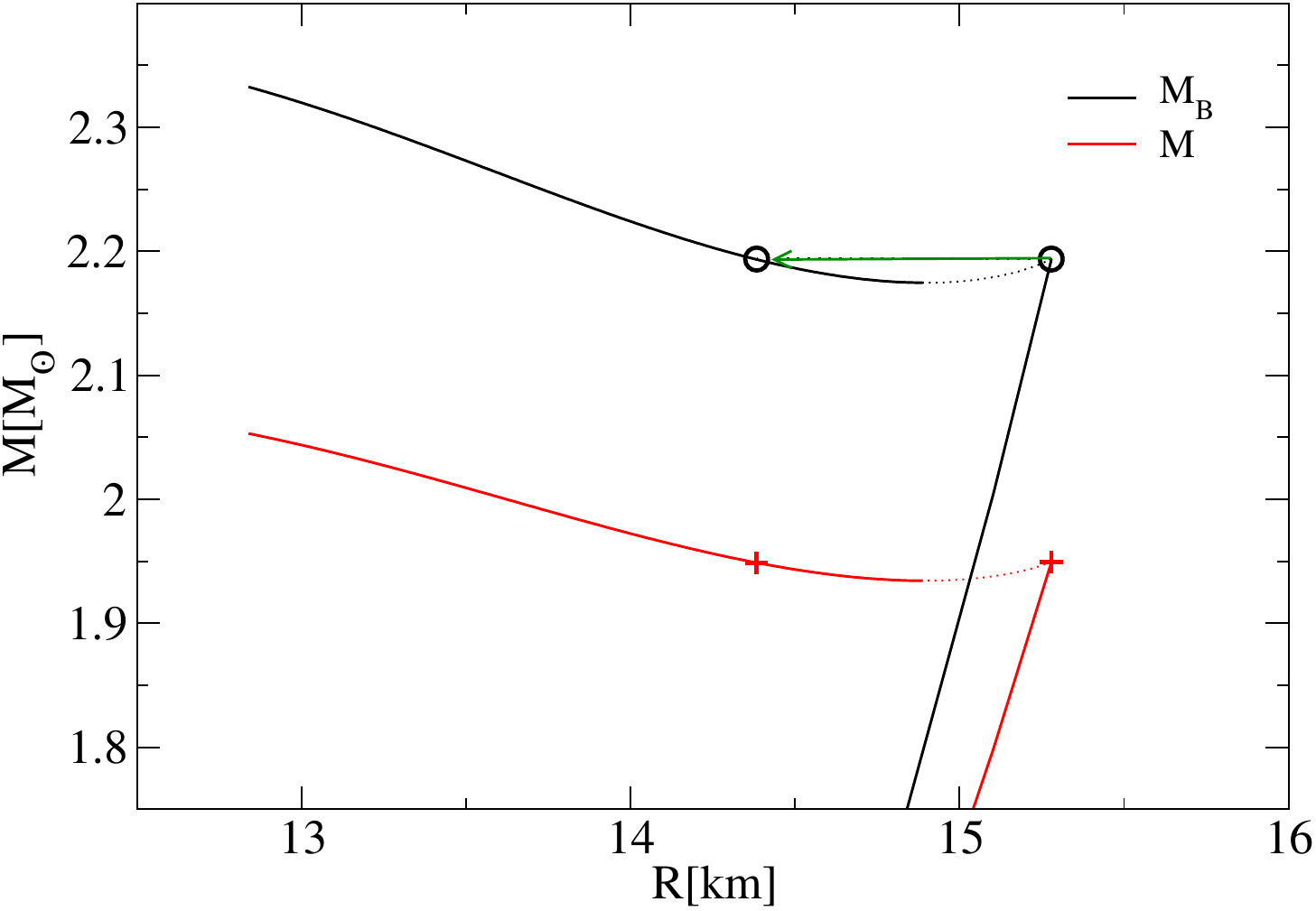}
\caption{\label{scheme-fig} Schematic representation of the neutron transition into a third branch. The black and red curves represent the baryonic and gravitational mass of static compact star sequences for a mass twin model. Dotted line represent unstable configurations against radial oscillations that do not fulfill the condition $ \partial M /\partial \varepsilon_c> 0$, see section~\ref{mrsection}. The green arrow line represents the trajectory of the transition under conservation of baryon number.}
\end{figure}

Within the scheme of this work it is assumed that the transition takes place between static, stable stellar configurations under conservation of the total baryon number. Figure~\ref{scheme-fig} shows a schematic plot of this scenario which includes baryonic and gravitational mass-radius curves. In this figure the green arrow indicates the path of the transition from the last stable stellar hadronic configuration into the hybrid one of smaller radius, preserving baryonic mass. For previous calculations under the same approach for energetics of the transition, see~(\cite{Alvarez-Castillo:2015dqa, Alvarez-Castillo:2016dyz}).
This article is organized as follows. In section 2 we introduce the equation of state and present our choice of parameter values. In section 3 we focus on the mass-radius relation diagram for neutron stars that includes allowed regions derived from measurements and various theoretically derived constrains as well as the stellar sequences for the chosen EoS sets. Section 3 presents the resulting available energy estimates for all our models. We finalize this manuscript with an outlook and discussion.

\section{The equation of state}

Detection of massive neutron stars of about 2M$_{\odot}$ has revealed that their EoS is rather stiff. For this reason we chose the density dependent functional DD2 EoS~(\cite{Typel:2009sy}) with excluded model correction~(\cite{Typel:2016srf}) that takes into account the finite size effect of nucleons due to the repulsive interaction between their internal quarks produced by Pauli blocking effects. The DD2 EoS correctly describes empirical data at sub-saturation and saturation densities and only varies at supra-saturation values where the behaviour of matter is uncertain. This is the domain where the excluded volume corrections hold. 

\begin{figure}
\center	
\includegraphics[width=70mm,height=50mm]{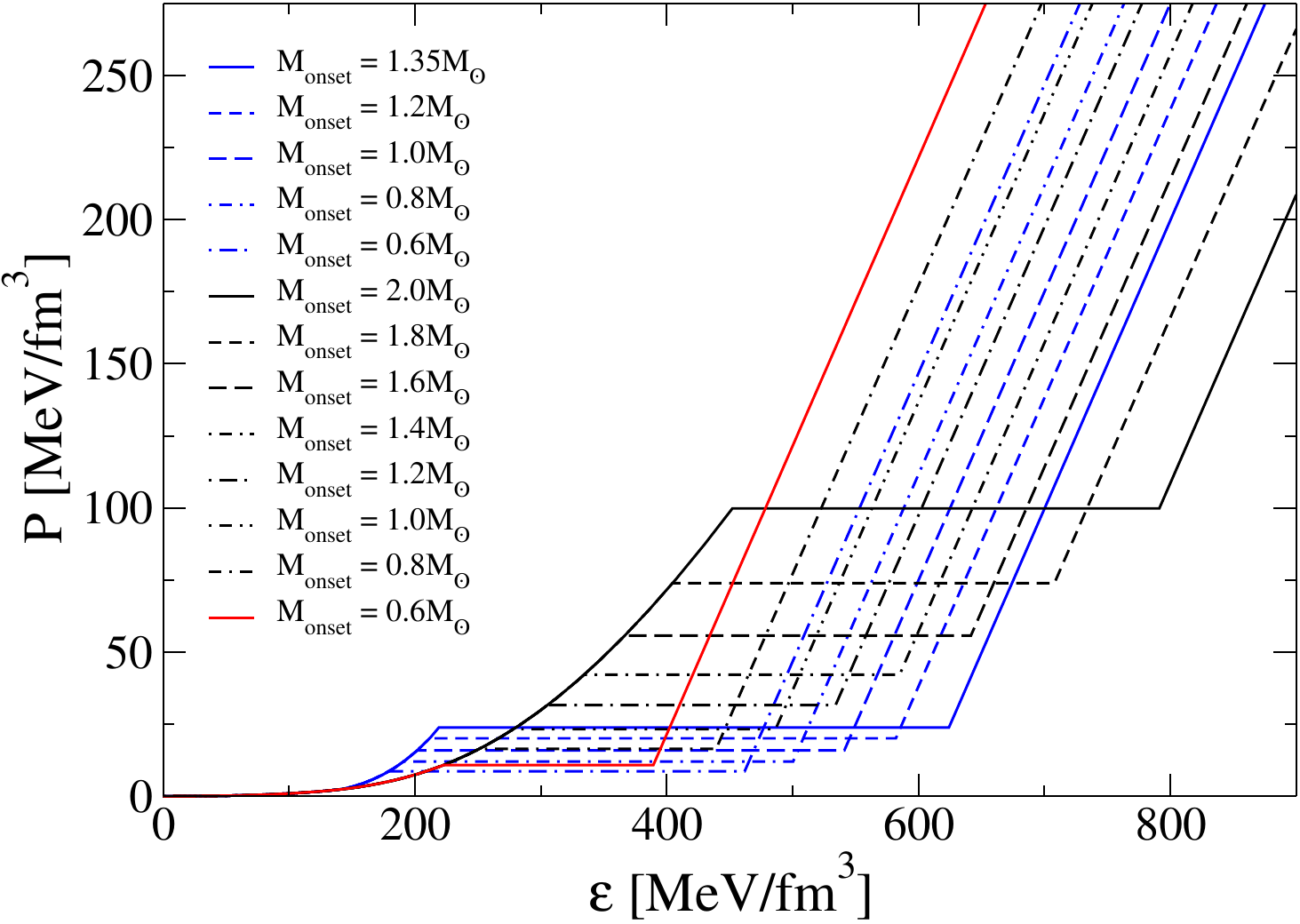}
\caption{\label{EoS-Fig} Equation of state diagram for set 1 (blue lines) and set 2 (black and red lines).}
\end{figure}

Additionally, the highest density part of the EoS consists of deconfined quark matter that is modelled with the constant speed of sound (CSS) formulation. Consequently, hadronic matter suffers a phase transition to quark matter via a Maxwell construction where their pressures equal. We follow the scheme introduced in~(\cite{Alford:2013aca}):
\beq
\ep(p) = \left\{\!
\begin{array}{ll}
\ep_{\rm NM}(p) & p<\ptrans \\
\ep_{\rm NM}(\ptrans)+\De\ep+c_{\rm QM}^{-2} (p-\ptrans) & p>\ptrans
\end{array}
\right.\ ,
\label{EoSqm}
\eeq
where $\varepsilon_{NM}$ refers to pure hadronic matter, $p_{trans}$ is the value of the pressure at the phase transition, $c_{\rm QM}$ is the speed of sound in quark matter and $\De\ep$ is the discontinuity in energy density at the transition. Moreover, the neutron star dense matter EoS described above is complemented with a well established neutron star crust EoS as given in~(\cite{Douchin:2000kad}).
An important relation to consider when discussing twin stars is the Seidov condition $\De\ep>\destab$~(\cite{Seidov:1971}) which ensures the appearance of a disconnected branch in the mass-radius relation of compact stars. The critical values at the transition are related by
\beq
\frac{\destab}{\etrans} = \frac{1}{2} + \frac{3}{2}\frac{\ptrans}{\etrans} \ .
\label{eqn:stability}
\eeq
In order to asses the maximum energy budget at the transition, we introduce two EoS sets with different excluded volume parameters for the hadronic DD2 EoS but sharing the same quark EoS with $c_{QM}=1$, described by Eq. (\ref{EoSqm}) . Set 1 corresponds to sequences with rather low mass onset $M_{onset}$, below 1.4M$_{\odot}$, whereas set 2 includes a wider range of values. Given the topology of the constraints in the mass-radius diagram, set 1 can consist of very stiff hadronic EoS only for those low mass onsets, see Fig.~\ref{MvsR-set1}. It will be shown in the results that the general tendency is that the lower the mass onset results in a higher maximum mass in the third branch. An important piece of information is that for set 1 the parameter $\gamma=\Delta\varepsilon/\varepsilon_{trans}$ has been varied in order to maximize the radius difference between their twin configurations while simultaneously fulfilling the constraints of~(\cite{Annala:2017llu,Bauswein:2017vtn}). On the contrary, for set 2 $\gamma=0.75$ remains constant in all the models, satisfying allowed radius values. Figure \ref{EoS-Fig} presents EoS curves for both set 1 and set 2 and shows that the the value of the critical pressure at the transition $p_{trans}$ is proportional to  the mass onset $M_{onset}$ for quark matter content in compact stars. The values of the EoS parameters are listed in Table \ref{table_parameters}.
\begin{table}[h]
\begin{center}
\resizebox{\columnwidth}{!}{%
\begin{tabular}{c|ccccc}
\hline \hline 
Model &	$M_{onset}$ &	 $n_{trans}$ & 	$\varepsilon_{trans}$ &$p_{trans}$  &$\Delta \varepsilon$\\
&					[M$_{\odot}$] &			[1/fm$^{3}$] &			[MeV/fm$^{3}$]  &		[MeV/fm$^{3}$]  & [MeV/fm$^{3}$]\\
\hline
DD2p80 &1.35 &0.225 &219.001 &23.836 & 354.782 \\
DD2p80 &1.2 &0.218 &211.646 &20.126 & 370.38\\
DD2p80 &1.0 &0.209 &201.886 &15.923 & 339.168\\
DD2p80 &0.8 &0.198 &190.987 &12.054 & 309.398\\
DD2p80 &0.6 &0.186 &179.03 &8.651 & 282.868\\
\hline
DD2p15 &2.0 &0.434 &452.144 &99.817 & 339.108\\
DD2p15 &1.8 &0.396 &404.544 &73.897 & 303.408\\
DD2p15 &1.6 &0.364 &366.622 &55.689 & 274.966\\
DD2p15 &1.4 &0.336 &334.285 &42.165 & 250.714\\
DD2p15 &1.2 &0.310 &305.229 &31.636 & 228.922\\
DD2p15 &1.0 &0.284 &277.994 &23.258 & 208.495\\
DD2p15 &0.8 &0.258 &251.022 &16.449 & 188.267\\
DD2p15& 0.6 &0.230 &222.456 &10.805 & 166.842\\
\hline
\end{tabular}
}
\end{center}
\caption{\label{table_parameters}EoS parameters for two sets: set 1 (DD2p80) and set 2 (DD2p15). See text for details.}
\end{table}

\section{The mass-radius diagram of compact stars}
\label{mrsection}

\begin{figure}
\center	
\includegraphics[width=70mm,height=70mm]{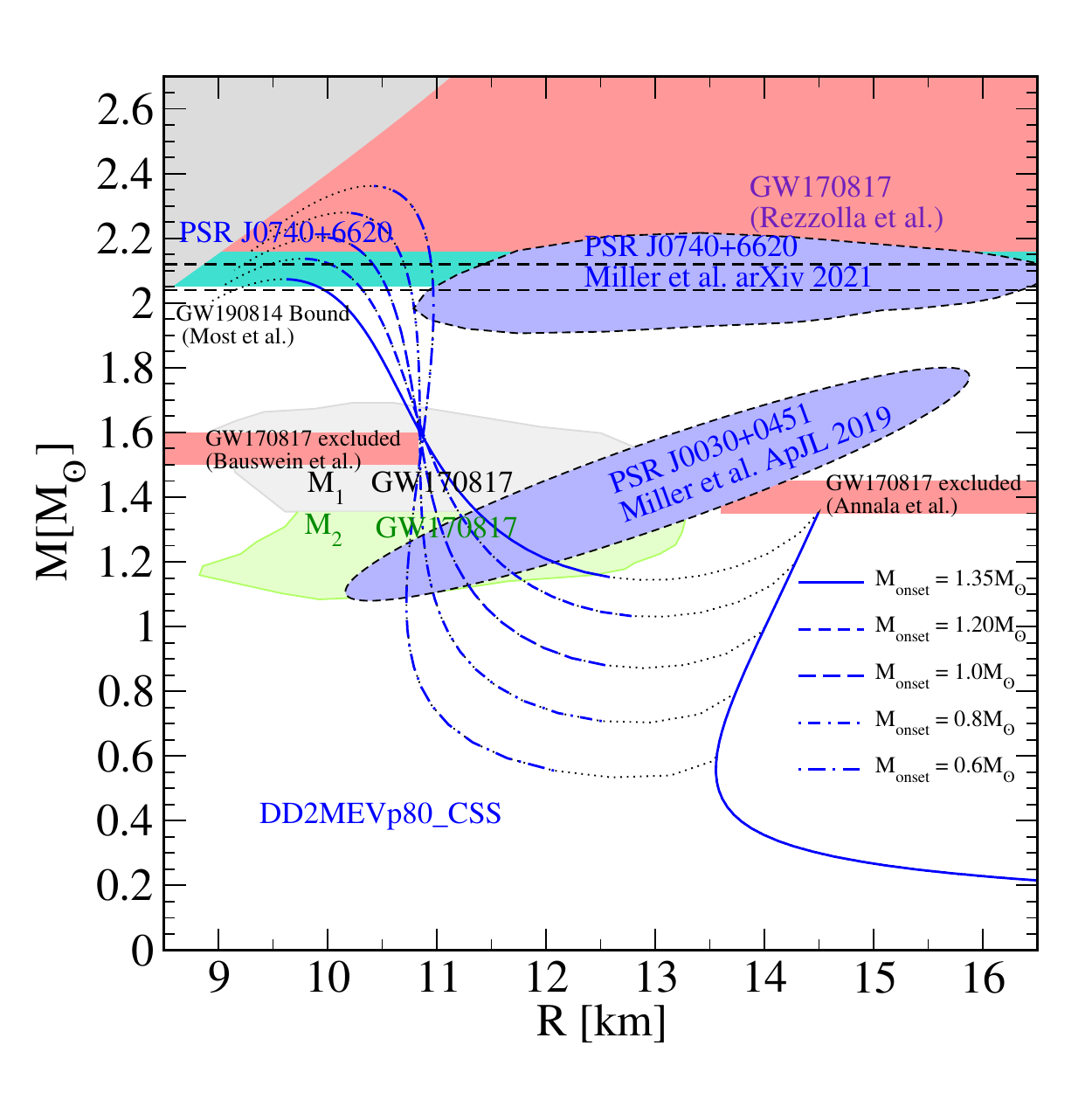}
\caption{\label{MvsR-set1} Mass radius diagram for the equations of state of set 1 consisting of the hadronic DD2p80 and quark CSS models, respectively. The resulting sequences feature low values of $M_{onset}$, with the property that the lower the mass onset, the higher maximum mass value in the third branch. Modern constraints from
multi-messenger astronomy are also displayed: mass measurement of PSR J0740+6620~(\cite{Cromartie:2019kug}), mass and radius measurements for
PSR J0030+0451 and PSR J0740+6620 by NICER as reported in~(\cite{Miller:2019nzo}) and in~(\cite{Miller:2021qha}) respectively, as well as and neutron star compactness from the tidal deformabilities measured in the compact star merger event GW170817, derived from the analysis the gravitational wave signal from the inspiral phase~(\cite{Abbott:2018exr}).
Red bands are forbidden regions derived also from the same analysis of the GW170817 signal by Bauswein et al.~(\cite{Bauswein:2017vtn}), Annala et al.~(\cite{Annala:2017llu}) and Rezzolla et al.~(\cite{Rezzolla_2018}). The latter one valid under the assumption that the compact stars merger collapsed into a black hole. The horizontal band delimited by dashed lines that is centred around 2.08M${_\odot}$ corresponds to a lower boundary on the maximum neutron star mass derived from the event GW190814 signal under the assumption that one of the components of the merger was a fast rotating neutron star~(\cite{Most:2020bba}). The upper left grey region is unaccessible to neutron stars due to causality violation on the equation of state.}
\end{figure}

\begin{figure}
\center	
\includegraphics[width=70mm,height=70mm]{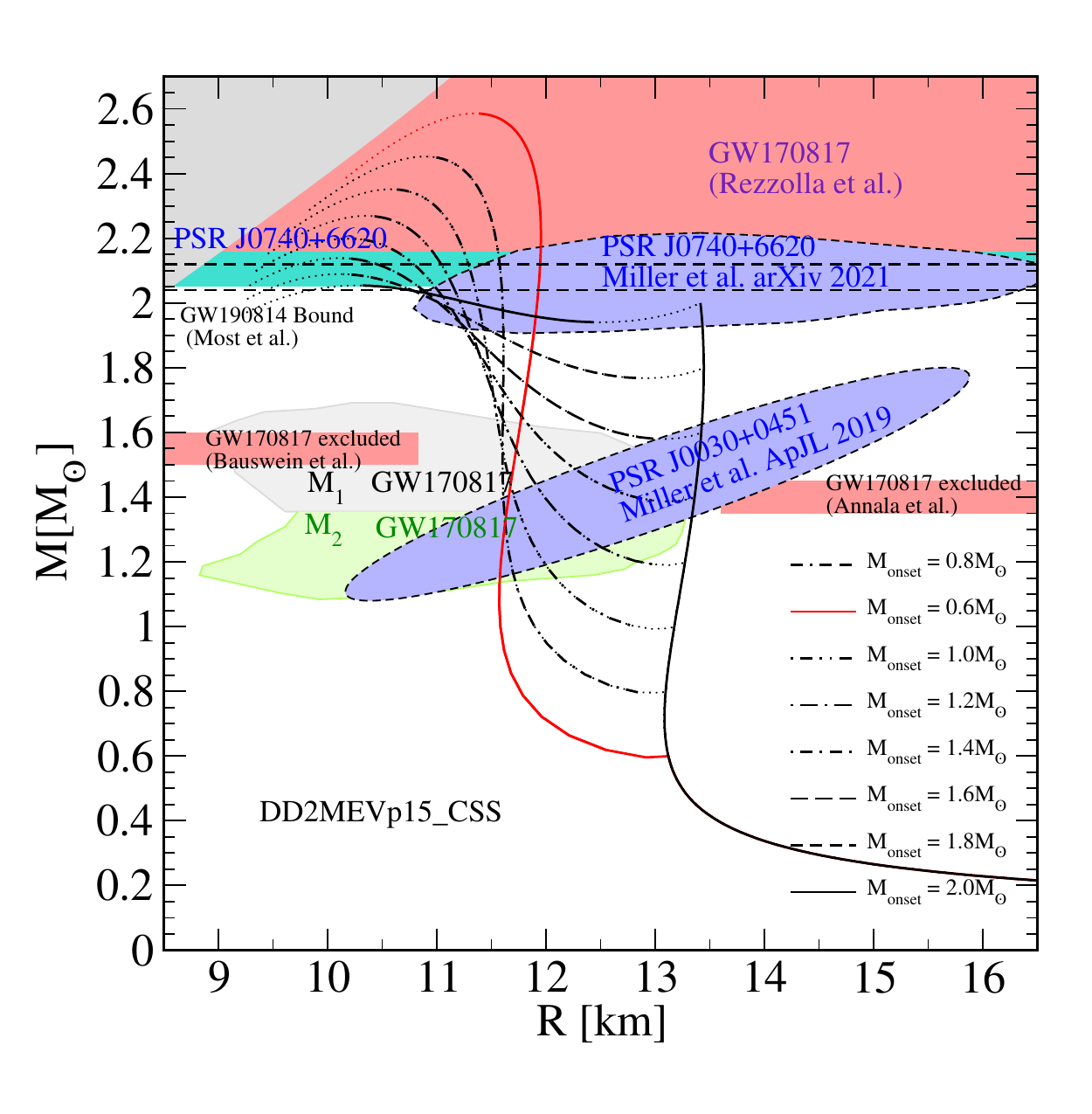}
\caption{\label{MvsR-set2} Mass radius diagram for the equations of state of set 2 consisting of the hadronic DD2p15 and quark CSS models, respectively. The resulting sequences feature a set of wider values of $M_{onset}$ than set 1 that fulfill the  same mass-radius constraints displayed in figure \ref{MvsR-set1}. The red curve for $M_{onset}=0.6M_{\odot}$ represents a single, connected branch where the equation of state lacks the sufficient latent heat $\Delta\varepsilon$ to feature unstable configurations thus not producing the mass twins scenario.}
\end{figure}

For an equation of state in the form of $p(\varepsilon)$, the compact star properties of static configurations are derived by solving the Tolman-Oppenheimer-Volkoff equations~(\cite{Tolman:1939jz,Oppenheimer:1939ne}):
\begin{eqnarray}
 \label{TOV}
\frac{\dd P( r)}{\dd r}&=& 
-\frac{\left(\varepsilon( r)+P( r)\right)
\left(m( r)+ 4\pi r^3 P( r)\right)}{r\left(r- 2m( r)\right)},\\
\frac{\dd m( r)}{\dd r}&=& 4\pi r^2 \varepsilon( r).
\label{eq:TOVb}
 \end{eqnarray}
The total mass of the star is defined by $M=m(r=R$) for the star radius $R$ where the pressure at the surface vanishes $p(r=R)=0$, and is complemented by the condition $m(r=0)=0$. The initial condition for the solution of the above equations is the determination of the energy density value at the center of the star, $\varepsilon_c$.
Sequences of compact stars are produced by solving for a star with a central energy density of about nuclear saturation and then
considering an increase of this value to be used as an input for another compact star, resulting in two points in the mass-radius diagram. This process is to repeated until the most massive compact star in the third branch is found. In figures \ref{MvsR-set1} and \ref{MvsR-set2} the mass-radius diagrams display entire sequences of compact stars as continuous lines with exceptions where unstable configurations appear, displayed as dotted lines to guide the eye. Those unstable compact star configurations fail to fulfill the stability condition against radial oscillations that is given by $ \partial M /\partial \varepsilon_c> 0$. Colourful regions in the mass radius diagram include state-of-the art constraints from multi-messenger astronomy observations as described in the figure \ref{MvsR-set1} caption. In particular, the derivations of a possible lower bound on the maximum compact star by~(\cite{Rezzolla_2018}) and by~(\cite{Most:2020bba}) have a strong dependence on the assumptions on the components and remnants of the mergers of GW170817 and GW190814, therefore we do not strictly rule out EoS models leading to maximum masses above 2.04M$_{\odot}$. Interestingly, recent X-ray observations of prolonged emissions of the remnant of GW170817 suggest the possibility of the formation of a compact star in lieu of a black hole~(\cite{Troja:2020pzf}).
\section{Results and Conclusions}

\begin{figure}
\center	
\includegraphics[width=70mm,height=70mm]{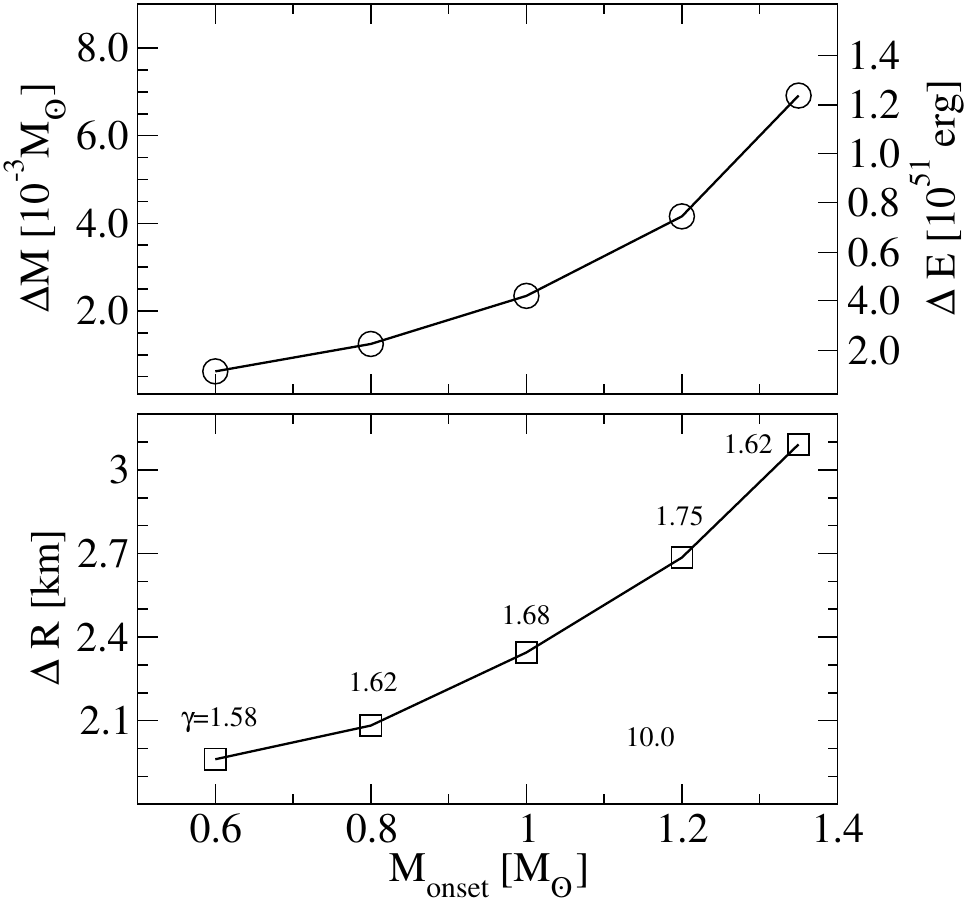}
\caption{\label{Energy-set1} Maximum energy disposable during a neutron star transition into a third family under conservation of baryon number. These results correspond to set 1, with variable $\gamma=\Delta\varepsilon/\varepsilon_{trans}$~ parameter.}
\end{figure}

\begin{figure}
\center	
\includegraphics[width=70mm,height=70mm]{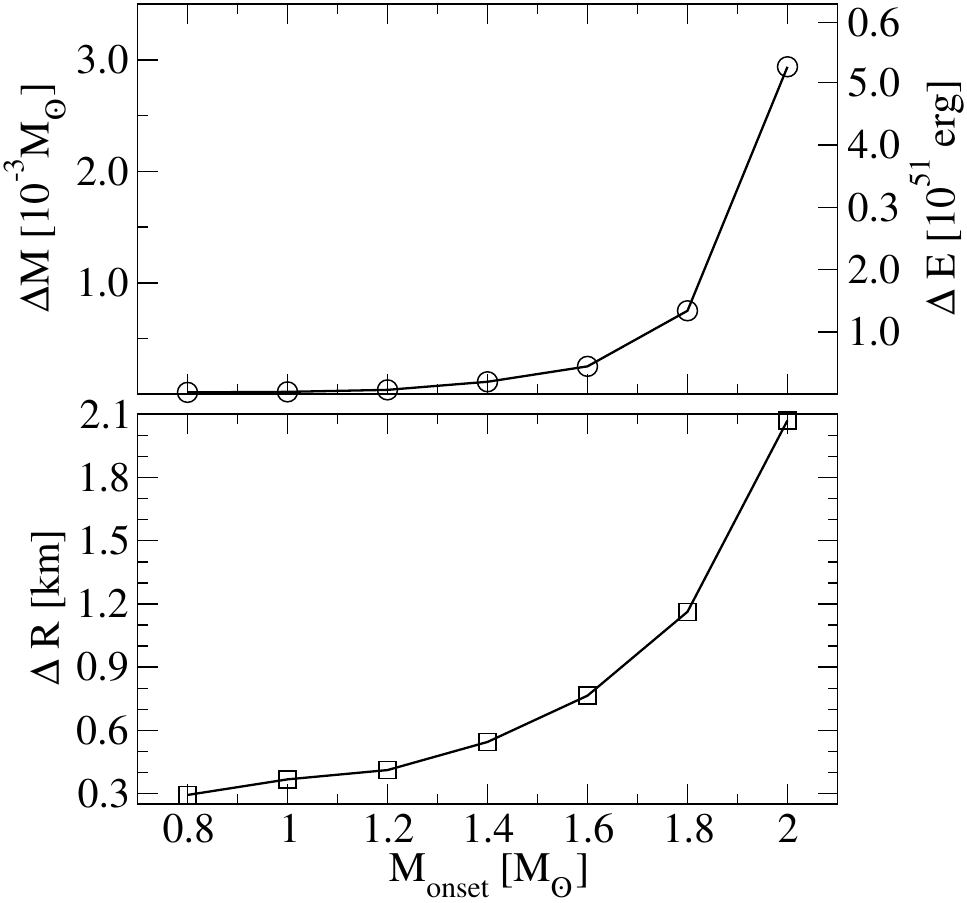}
\caption{\label{Energy-set2} Maximum energy disposable during a neutron star transition into a third family under conservation of baryon number. These results correspond to set 2, where the $\gamma=\Delta\varepsilon/\varepsilon_{trans}$~ parameter is fixed.}
\end{figure}

In this study we have prepared two EoS sets with different stiffness of hadronic matter but sharing the same maximal stiff quark matter leading to different compact star mass onsets $M_{onset}$ for quark deconfinement, their sequences satisfying modern constraints from multi-messenger astronomy.  Moreover, we computed the energy released at the transition from $M_{onset}$ into the third branch under conservation of baryonic number. This energy values are, by construction, maximal for each of the EoS models. Figures \ref{Energy-set1} and \ref{Energy-set2} show  the energy released values and radius difference of compact stars  as a function of $M_{onset}$ following the stellar transition for set 1 and set 2, respectively. We find an opposite tendency in our results from what has been reported in~(\cite{Alvarez-Castillo:2015dqa, Alvarez-Castillo:2016dyz}), where both the energy release $\Delta E$ and radius difference $\Delta R$ decrease as a function of $M_{onset}$. Comparing the transitions properties for set 1 and set 2 we find that the available energy values can be two orders of magnitude different for some of the model parameters, with set 1 displaying the highest values. Interestingly, the existence of set 1-like models for a similar EoS~(\cite{Alvarez-Castillo:2018pve}) with a low $M_{onset}$ have been favoured by a recent Bayesian analysis that provides posterior probabilities based on the capability of a given model to simultaneously fulfill exactly the same constrains we have considered here and additionally introduces a fictitious mass-radius measurement of the mass twins~(\cite{Blaschke:2020qqj}).
The extreme compactness of hybrid stars brings the possibility of  unambiguous identification  when special mass-radius or tidal deformabilities~(\cite{De:2018uhw}) measurements be performed.  A more exhaustive study on the mass-radius configurations for our EoS of choice has revealed that there exists exclusive mass-radius diagram regions where hybrid stars can be located, unreachable for pure hadronic or hyperonic configurations, namely small massive stars of about 2M$_{\odot}$ for which $R_{2.0}$ < 12km holds~(\cite{Cierniak:2020eyh}).  The argumentation is based on  realistic, hadronic EoS calculations that consider baryon-baryon two-body and three-body interaction featuring a repulsive short-range multi-pomeron exchange potential, resulting in larger compact stars~(\cite{Yamamoto:2015lwa,Yamamoto:2017wre}).

It has been shown that this type of energetic events might have an imprint on the properties of the trajectories of binary systems, whose large eccentricities can be explained by a pulsar kick following the transition without disrupting the system~(\cite{Alvarez-Castillo:2019apz}). It is also expected that this transition occurs in the evolution of the remnant of a compact star merger during the process of formation of a more massive hybrid star or a black hole. Future multi-messenger observations are expected to provide much more precise information on these kind of catastrophic events.

In addition, energetic emissions from compact stars play an important role in the understanding of the nature of cosmic rays and their implications for fundamental physics and cosmology. Therefore, we hope that the estimates presented in this work help for this purpose. For a review on several aspects of cosmic rays studies, see~(\cite{Homola:2020odt}).

\section{Acknowledgments}
The author would like to thank Stefan Typel for providing the hadronic equation of state tables for this study and acknowledges support from the the Bogoliubov-Infeld program for collaboration between JINR and Polish Institutions as well as from the COST actions CA15213 (THOR) and CA16214 (PHAROS).


\vspace{-1.75cm}
\end{document}